\documentclass[aps,pre,twocolumn,groupedaddress,showpacs]{revtex4}
\usepackage{graphicx}
\usepackage{amssymb}

\begin{document}
\title{Geometry dependence of the clogging transition in tilted hoppers}
\author{C. C. Thomas and D. J. Durian}
\affiliation{Department of Physics and Astronomy, University of Pennsylvania, Philadelphia, PA 19104-6396, USA} 

\date{\today}

\begin{abstract}
We report the effects of system geometry on the clogging of granular material flowing out of flat-bottomed hoppers with variable aperture size $D$, and with variable angle $\theta$ of tilt of the hopper away from horizontal.  In general, larger tilt angles make the system more susceptible to clogging.  To quantify this effect for a given $\theta$, we measure the distribution of mass discharged between clogging events as a function of aperture size and extrapolate to the critical size at which the average mass diverges.  By repeating for different angles, we map out a clogging phase diagram as a function of $D$ and $\theta$ that demarcates the regimes of free flow  (large $D$, small $\theta$) and clogging (small $D$, large $\theta$).  We do this for both circular holes and long rectangular slits.  Additionally, we measure four types of grain: smooth spheres (glass beads), compact angular grains (beach sand), disk-like grains (lentils), and rod-like grains (rice). For circular apertures, the clogging phase diagram is found to be the same for all grain types.  For narrow slit apertures and compact grains, the shape is also the same as for circular holes when expressed in terms of projected area of the aperture against the average flow direction.  For lentils and rice discharged from slits, the behavior is different and may be due to alignment between grain and slit axes.
\end{abstract}

%
\maketitle






\section{Introduction}

Granular media are non-equilibrium, athermal systems that continue to evade a comprehensive physical description \cite{JaegerReview, DuranBook,CMMP2010}. One of the more intriguing properties of granular flow is the phenomenon of clogging, where the flow is spontaneously arrested due to obstacles or boundaries \cite{To01,To02,HelbingPRL,Chevoir07,Evesque07,Geiger07,Garcimartin10,EField,Obstacle,Drescher}.  This can be seen in everyday life, whether as salt clogging in a shaker, or pedestrian traffic at a theater exit \cite{HelbingNature}. Understanding and controlling clogging is important for industry and agricultural processing. In most applications, the goal is typically either to eliminate clogging from a system entirely, or else to use clogging to control output volume, such as metering a bit of salt from a shaker. Furthermore, clogging, together with fundamental questions about the nature of the glass transition, has been responsible for motivating much of the rich field of jamming (for reviews, see \cite{LiuNagelBook, LiuNagelChapter}).

While most work on jamming focuses on spatially uniform systems, clogging behavior is determined by convergent flow toward an opening in the boundary. The sudden formation of a stable arch over the exit halts all flow, and then the entire system is then jammed. Other systems which demonstrate clogging include the flow of vortices through an array of pinning sites in type-II superconductors as well as grains flowing through an array of obstacles \cite{Reichhardt,Reichhardt2012,GoodrichBug}. These systems demonstrate a clogging transition which depends on the density of pinning sites and the packing fraction of particles.

As a canonical example of clogging, we consider the discharge of grains from the circular hole of diameter $D$ located on the floor of a flat-bottomed hopper or silo.  For three-dimensional hoppers, it has been observed that the time scale one must wait for grains to discharge before a clog occurs diverges at a critical aperture size $D_c$ \cite{Zuriguel05,Mankoc09}.  We understand this as a clogging transition.  For large apertures, $D>D_c$, the granular flow will never  clog, while for the small apertures, $D < D_c$, the flow proceeds steadily for a while but then eventually clogs. The location of the clogging transition is therefore found at $D=D_c$. Zuriguel et al.~\cite{Zuriguel05} found the value of $D_c$ by measuring the average mass of grains discharged before a clog occurs, $\langle m\rangle$, as a function of the exit hole diameter $D$. The value of $D$ at which $\langle m\rangle$ diverges is therefore the critical hole diameter $D_c$. They found for spherical grains that $D_c$ scales with the grain diameter $d$, as $D_c/d = 4.94 \pm 0.03$. In their work, they did not see a dependence of $D_c/d$ with either grain surface roughness or polydispersity. However, they did observe that $D_c/d$ depends on the material shape. Additional work found that vibrating the entire hopper reduced $D_c$ \cite{Mankoc09}. These studies on the clogging transition for three-dimensional hoppers stand in contrast to similar work with two-dimensional hoppers, where it was unclear whether the time to wait until a clog occurs diverged at a finite hole size~\cite{To05,Janda2D08,Tang09,Tang12}. For such geometries, the clogging behavior can be equally well described by a diverging time scale~\cite{To05,Janda2D08} and an exponential~\cite{Tang09,Tang12} or exponential squared~\cite{To05,Janda2D08} functions of $D$.

The formation of arches over a hole is affected by not only grain and hole sizes and shapes, but also by the angle $\theta$ at which the plane of the hole is tilted away from horizontal.  Tilting thus offers an alternative geometrical parameter that may be easily and continuously varied in order to affect clogging behavior.  This was explored by Sheldon and Durian \cite{Sheldon}, who measured both the flux of discharging grains and the clogging behavior as a function of hole size and tilt angle.  In analogy to the start and stop heights for granular flow down inclined planes \cite{Pouliquen}, they mapped out a ``clogging phase diagram" by continuously tilting the hopper and measuring the angles at which flow spontaneously stops (for increasing tilt angle) and starts (for decreasing tilt angle).  Since this gave two critical angles for a given hole size, and since both could depend on tilt rate, the clogging transition was not precisely located.  Furthermore, the connection between these start and stop angles and the critical aperture sizes of Refs.~\cite{Zuriguel05,To05,Janda2D08,Mankoc09} was not explored.

In this paper we map out the clogging phase diagram, determining a well-defined clogging transition in $D - \theta$ parameter space, for various hopper and grain geometries. We do this by observing the values of $D$ where there is a divergence in the mass of material discharged between clogging events. This is similar to the method used in prior experiments, which were only performed for circular holes and at $\theta = 0$ ~\cite{Zuriguel05, Mankoc09}. The value of $D_c$ found in those experiments describes the location of the transition at $\theta = 0$.  We find $D_c(\theta=0$) to be in agreement with those values.  We measure the variation of $D_c$ with $\theta$, and discover that the location of this transition is in agreement with the preliminary measure given by the start angles of Ref.~\cite{Sheldon}. We also map out the clogging transition for a variety of material shapes and sizes (see Table \ref{Materials}), and two different aperture shapes: a circular hole and a rectangular narrow slit.  Remarkably, for circular holes, we find that the shape of the clogging phase diagram is identical for all materials and that the shape for holes and slits is the same for compact grains.

\section{Experimental Details}

We study clogging for a variety of granular materials: Potters 2~mm and A100 series glass spheres, U. S. Silica Ottawa sand, lentils, and Basmati rice.  Table~\ref{Materials} lists the relevant properties of these grains, including the draining angle of repose $\theta_r$, the bulk density of the granular packing $\rho_b$, and the long $d$ and short $d_{short}$ axes of the grains. The maximum tilt angle, beyond which the medium loses contact with the wall and no grains can exit, is $\theta_{max}=\pi-\theta_r$. We measure $\theta_r$ by pouring the granular materials into a cylindrical container with a large circular hole in the center of the flat bottom. After the discharge comes to completion, $\theta_r$ is measured as the angle which the surface of the remaining grains makes with the horizontal.

\begin{table*}
\caption{\label{tab:Materials}Material properties and clogging results for the grains tested.  The long and short axes are denoted $d$ and $d_{short}$, respectively. The bulk density is $\rho_b$ and the draining angle of repose is $\theta_r$.  The value $k$ is the dimensionless hole size where the Beverloo equation (Eq.~(\ref{Beverloo})) predicts the flux to vanish.  The dimensionless critical aperture sizes at zero tilt, $D_{co}/d$, for both the circular hole and rectangular slit are also shown below.  Note that for all materials $D_{co}/d > k$. Error bars for $d$ and $d_{short}$ indicate the standard deviations of the distributions. For the other material properties, error bars indicate the standard error in the measurements.}
\begin{ruledtabular}
\begin{tabular}{llllllllll}
	Material & $d$ (mm) & $d_{short}$ (mm) & $\rho_b$ (g/cm$^3$) & $\theta_{r}$ & $k$ &  Hole $D_{co}/d$ & Slit $D_{co}/d$\\ \hline
	Glass Spheres & $2.02 \pm 0.04$ & ---   & $1.62 \pm 0.01$ & $23.5\pm 0.5^{\circ}$ & $1.5 \pm 0.1$ &  $4.5 \pm 0.2$ &---\\
	Glass Spheres & $0.96 \pm 0.05$  & ---  &$1.56 \pm 0.02$ & $20.0\pm0.4^{\circ}$ & --- & --- &$1.68 \pm 0.09$\\
	Ottawa Sand & $0.77 \pm 0.10$   & ---  &$1.61 \pm 0.06$& $33.6\pm0.5^{\circ}$ & $1.7 \pm 0.6$  &$6.1\pm0.8$&$2.1\pm0.3$\\
	Basmati Rice & $7.4 \pm 0.5$ & $1.48 \pm 0.10$   & $0.81 \pm 0.04$ & $34.2\pm0.7^{\circ}$ & $0.8 \pm 0.2$ & $2.5 \pm 0.2$ &$1.31 \pm 0.09$\\
	Lentils &  $5.7 \pm 0.5$ & $2.3 \pm 0.2$ & $0.83 \pm 0.02$  & $30.2\pm0.6^{\circ}$ &   $1.0 \pm 0.4$ &$2.8 \pm 0.2$ &$1.5 \pm 0.1$\
\label{Materials}
\end{tabular}
\end{ruledtabular}
\end{table*}

\begin{figure}
\includegraphics[width=3 in]{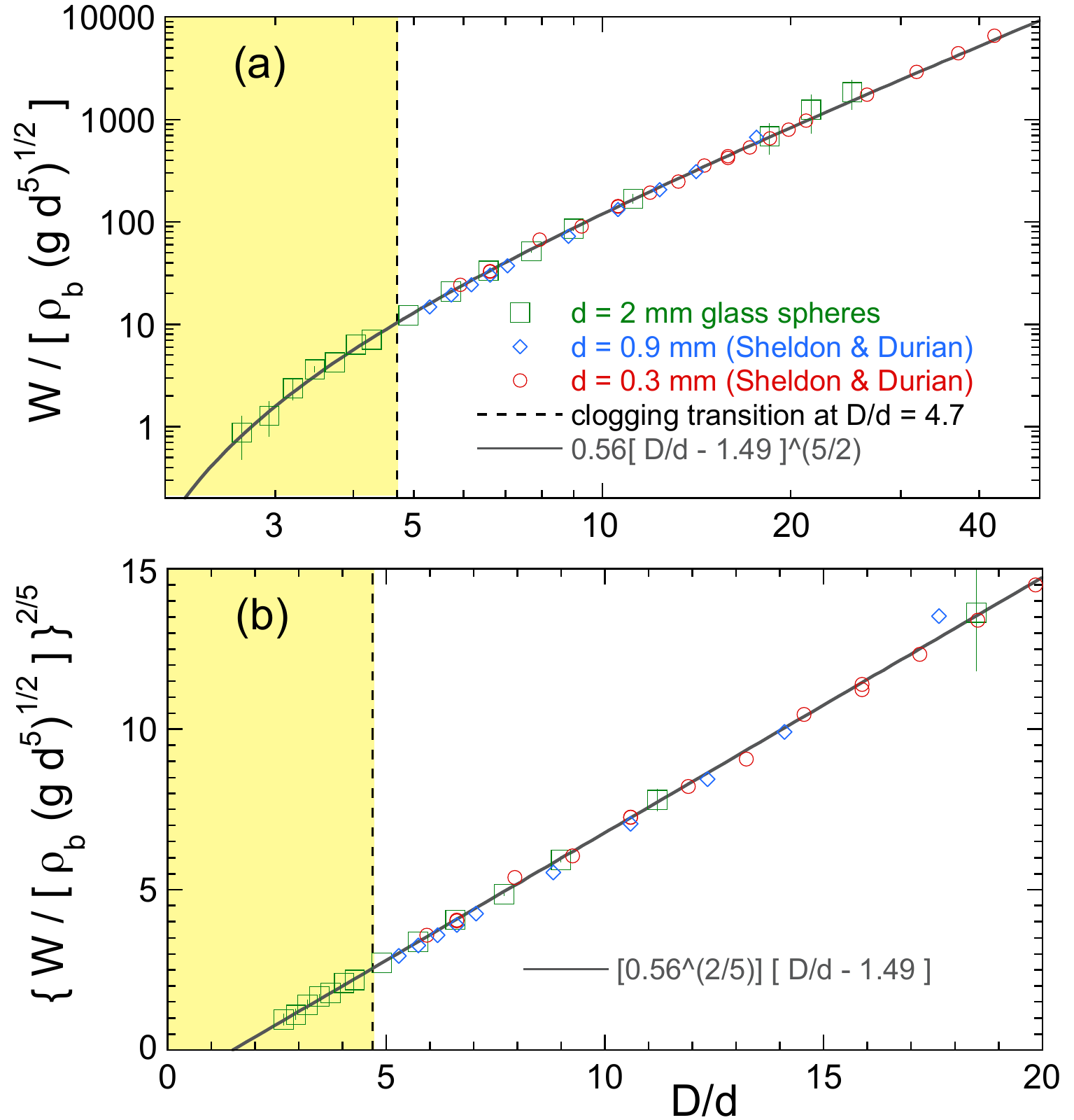}
\caption{(Color online) (a) Dimensionless discharge rate $W/(\rho_b \sqrt{ g d^5})$, and (b) the same quantity raised to the 2/5 power, for glass beads as a function of dimensionless hole diameter $D/d$, where $W$ is the mass discharge rate, $\rho_b$ is the bulk density of the packing, $g=9.8$~m/s$^2$, and $d$ is the grain diameter as labeled.  The data for the $d=2$~mm spheres are new, while the other data are taken from Ref.~\cite{Sheldon}.  The fit to the Beverloo form, Eqs.~(\ref{Beverloo},\ref{Bev2}), gives $C=0.56$ and $k=1.5$, and is shown a solid gray curve (a) and line (b).  The shaded region to the left of $D/d=4.7$ indicates where the system is susceptible to clogging}
\label{WScaling}
\end{figure}

To further characterize the grains, we also measure the mass discharge rate $W$ versus hole diameter $D$.  The results are well-described by the usual Beverloo equation,
\begin{eqnarray}
       W &=& C \rho_b\sqrt{g}(D-kd)^{5/2}, \label{Beverloo} \\
            &=& C [\rho_b\sqrt{gd^5}](D/d-k)^{5/2}, \label{Bev2}
\end{eqnarray}
where $C$ and $k$ are dimensionless fitting parameters, and $g=9.8$~m/s$^2$ \cite{Beverloo,NeddermanSavage}.  For the $d=2$~mm glass spheres, discharge rate data are divided by $\rho_b \sqrt{g d^5}$ and plotted versus $D/d$ in Fig.~\ref{WScaling}, along with prior data for two other sphere sizes \cite{Sheldon}.  All three data sets then collapse, as expected from Eq.~(\ref{Bev2}).  The  simultaneous fit to this equation is very good, as seen in the log-log plot of Fig.~\ref{WScaling}a as well as in the linear plot of Fig.~\ref{WScaling}b, where the dimensionless discharge rate is raised to the $2/5$ power thus making the Beverloo form a straight line that vanishes at $D/d=k$.  The fitted value of $k$ for the glass spheres is $1.5 \pm 0.1$, which is about three times smaller than the dimensionless critical hole size $D_c/d=4.7$ at which we find in later sections for the location of the clogging transition.  Note in Fig.~\ref{WScaling} that the discharge data is smooth and continuous, in accord with the Beverloo equation, on both sides of the clogging transition.  One might have expected a discontinuity in rate or slope at the transition, but this same continuous behavior is also observed for the other grain types.  Fitted values of $k$ are collected in Table~\ref{Materials}.

We investigate clogging through two aperture shapes: circular holes and narrow rectangular slits. For circular holes we mount camera irises of adjustable diameter $D$ on the bottom or sidewall of a rectangular aluminum hopper with inner dimensions $9.5\times9.5$~cm$^2$ and height 91~cm.  For $\theta<60^\circ$, we use the iris centered at the bottom of the hopper.  For $\theta>60^\circ$, we instead use the iris mounted on the sidewall at 5.0~cm above the bottom.  The location of the aperture at various locations along the bottom and sidewall was shown in Ref.~\cite{Sheldon} to have no effect on the rate of discharge, provided that the top free surface of the granular medium is many hole diameters aways from the aperture.  Intuitively, this is because the discharge rate is not set by a hydrostatic pressure but rather by the free-fall of grains from a broken transient arch.  We confirmed that, similarly, the location of the aperture on bottom versus sidewall does not affect the location of the clogging transition. This can also be seen by the absence of any discontinuity or kink in the clogging transition curve at $\theta=60^\circ$ (shown later, in Fig.~\ref{AllTransitions}).

We also investigate clogging for narrow rectangular slits of constant length 149~mm. For this, a custom-made slit of adjustable width $D$ is mounted on the bottom or side of a flat-bottomed container.  We bevelled the edges on the outside of the slit to ensure that the aperture wall thickness does not affect the clogging behavior. For $\theta<60^\circ$ we use a hopper with inner cross-section $28\times20$~cm$^2$, height 23~cm, and with the slit mounted on the bottom of the hopper.  For $\theta>60^\circ$, we use a hopper with inner cross-section $22\times14$~cm$^2$, height 30~cm, and with the slit mounted on the sidewall of the hopper.

The clogging behavior of all four classes of materials was tested for both the hole and the slit.  However, the 2~mm glass spheres were only used with the circular hole and the A100 glass spheres were only used with the slit. For the 2~mm spheres, we refilled the hopper during discharge.  For the other materials, we did not refill the hopper during flow, and restricted our maximum time window to the time to nearly exhaust the entire hopper.  In all cases, we ensured that the height of the grains at the end of discharge was large compared to $D$.  In such a limit, the discharge behavior is independent of the filling height.  Experiments~\cite{NeddermanSavage} and simulations~\cite{Anand08} confirm that flux is generally independent of filling height as long as the height is larger than the hole size.

\section{Discharge Distributions}

\begin{figure}
\includegraphics[width=3 in]{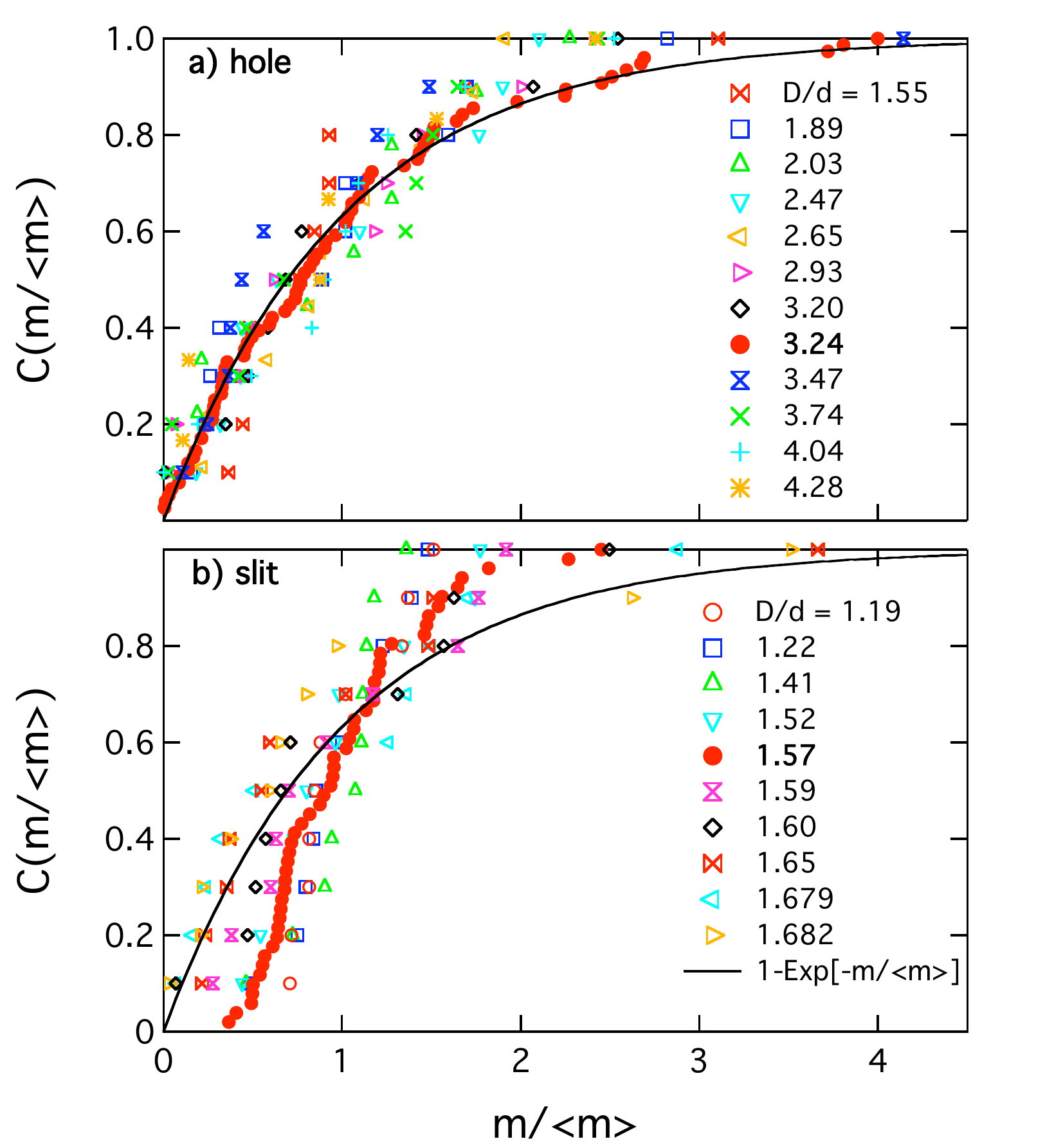}
\caption{(Color online) Cumulative distribution function $C$ of mass discharged $m$, normalized by the average mass $\langle m\rangle$ for a variety of hole diameters and slit widths $D$ as labeled.  Shown is $C(m/\langle m\rangle)$ for (a) $d$ = 2~mm diameter glass spheres discharged from a circular hole and (b) $d$ = 1~mm diameter glass spheres from a rectangular slit, at zero tilt.  The overlaid curve is the cumulative distribution function for an exponential distribution.  The distribution for the circular holes is similar to an exponential; for the rectangular slit it is somewhat sharper than an exponential.  These distributions are typical for all angles and materials studied.}
\label{CDF}
\end{figure}

\begin{figure}
\includegraphics[width=3in]{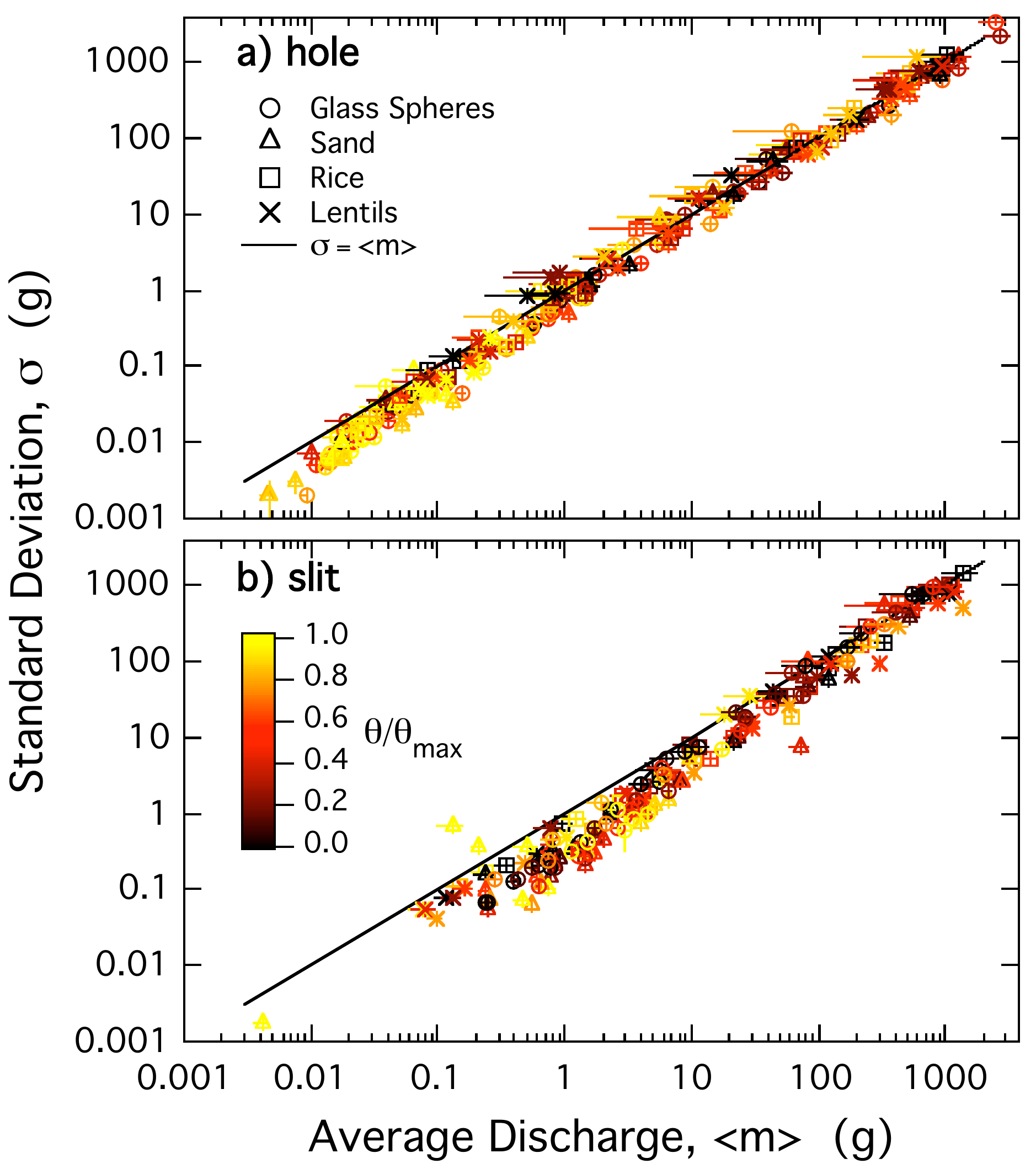}
\caption{(Color online) Standard deviation $\sigma$ of the discharged mass as a function of the average discharged mass $\langle m\rangle$ for the (a) circular hole and (b) rectangular slit.  Symbol shapes denote different grain types, with shading indicating the tilt angle $\theta / \theta_{max}$.  All aperture sizes $D$ are represented on the plots: for any given grain type and $\theta/\theta_{max}$, $\langle m \rangle$  increases with hole diameter or slit width $D$. The standard deviation is roughly proportional to the average.  The expected relationship for an exponential distribution is $\sigma = \langle m \rangle$, overlaid.  The discharge distributions through a slit are somewhat sharper than for an exponential distribution.}
\label{stdev}
\end{figure}

As a prelude to measuring the clogging transition by the method of Refs.~\cite{Zuriguel05,To05,Janda2D08,Mankoc09}, now as a function of tilt angle, we must first characterize the distribution of the mass discharged for a given hopper and grain geometry.  To initiate flow from a clogged configuration, we break the arch over the exit by gently poking it either with a wire (for holes) or stiff paper (for slits).  By repeating ten or more times, we acquire statistics for the cumulative distribution, $C(m)$, which is the fraction of events with mass less than or equal to $m$.  Fig.~\ref{CDF} illustrates typical behavior for glass spheres discharged at zero tilt angles through (a) holes and (b) slits of different sizes.  After scaling by the average discharged mass $\langle m\rangle$, the data for holes collapse nicely to a rising exponential $C(m/\langle m\rangle)=1-\exp(-m/\langle m\rangle)$, exhibiting no trend in hole diameter $D$.  The discharge distribution, which equals ${\rm d}C(m)/{\rm d}m$, is therefore exponential and it follows that the clogging is a random Poisson process that occurs with some rate, independent of prior history.  This agrees with previous observations \cite{Tang09, Tang12,Janda2D08, Zuriguel05, Mankoc09, Zuriguel03, To05,Janda09}.  For slits, in Fig.~\ref{CDF}b, the data do not collapse as nicely as for holes but yet display no particular trend with slit width.  The rise from zero to one is somewhat faster than exponential.  This differs from a previous experiment for a wedge hopper with a long narrow slit \cite{Saraf}, where flow was initiated by jolting the hopper and where a power-law distribution was observed.

The standard deviation $\sigma$ of the discharged masses is a simple parameter for characterizing the width of the distribution.  Results for $\sigma$ for ${\it all}$ grain types and tilt angles are plotted versus $\langle m\rangle$ in Fig.~\ref{stdev} for (a) holes and (b) slits.  For holes, the data span many orders of magnitude and closely agree with $\sigma=\langle m\rangle$, which is the expectation for a Poisson process with an exponential distribution.  For slits, $\sigma$ approach $\langle m\rangle$ for large $\langle m\rangle$, i.e.\ for large slit widths.  For smaller slit widths, $\sigma < \langle m\rangle$ holds and hence the discharge distribution is {\it sharper} than exponential, as seen explicitly for the glass beads at $\theta = 0$ in Fig.~\ref{CDF}b.  In all cases, we thus find that the distribution of discharged masses is exponential or sharper.  Therefore the average mass $\langle m\rangle$ can be determined to within $\pm 30\%$ uncertainty from only ten trials and used, next, to locate the clogging transition.

\section{Clogging Transition}

\begin{figure}
\includegraphics[width=3in]{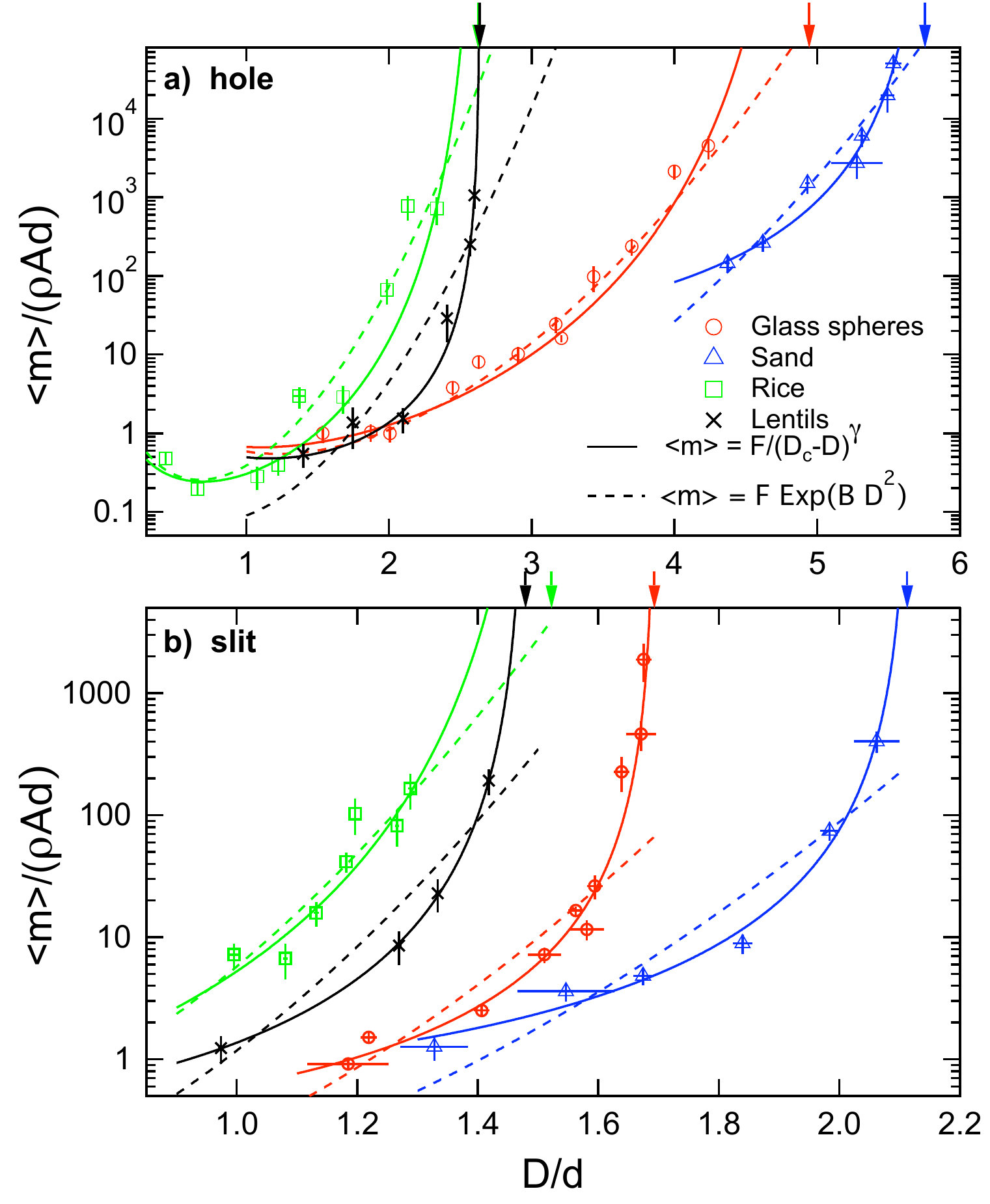}
\caption{(Color online) Dimensionless measure of the average mass discharged $\langle m\rangle/(\rho_b A d)$ before a clog occurs versus dimensionless aperture size $D/d$. $A$ is the aperture area, $\rho_b$ is the material bulk density, and $d$ is the long grain axis. For the circular hole, $D$ is the hole diameter, while for the rectangular slit, $D$ is the width of the slit. Plotted are data for (a) the circular hole and (b) the rectangular slit, for $\theta = 0$. The average mass grows with $D$ faster than exponential or even exponential squared (indicated by the dashed curves).  Instead, $\langle m\rangle$ versus $D$ is better described by the power-law of Eq.~(\ref{PowerLaw}), shown by the solid curves that diverge at critical aperture sizes $D_c$, indicated by the arrows.  The exponents, $\gamma$, were adjusted as part of the fits and the values are given in Fig.~\ref{Power}.  For the rice data, however, such fits did not converge and the exponents for the displayed fits were instead fixed to $\gamma=5$ in (a) and $\gamma=2$ in (b).}
\label{MassDZeroDeg}
\end{figure}

\begin{figure*}
\includegraphics[width=6in]{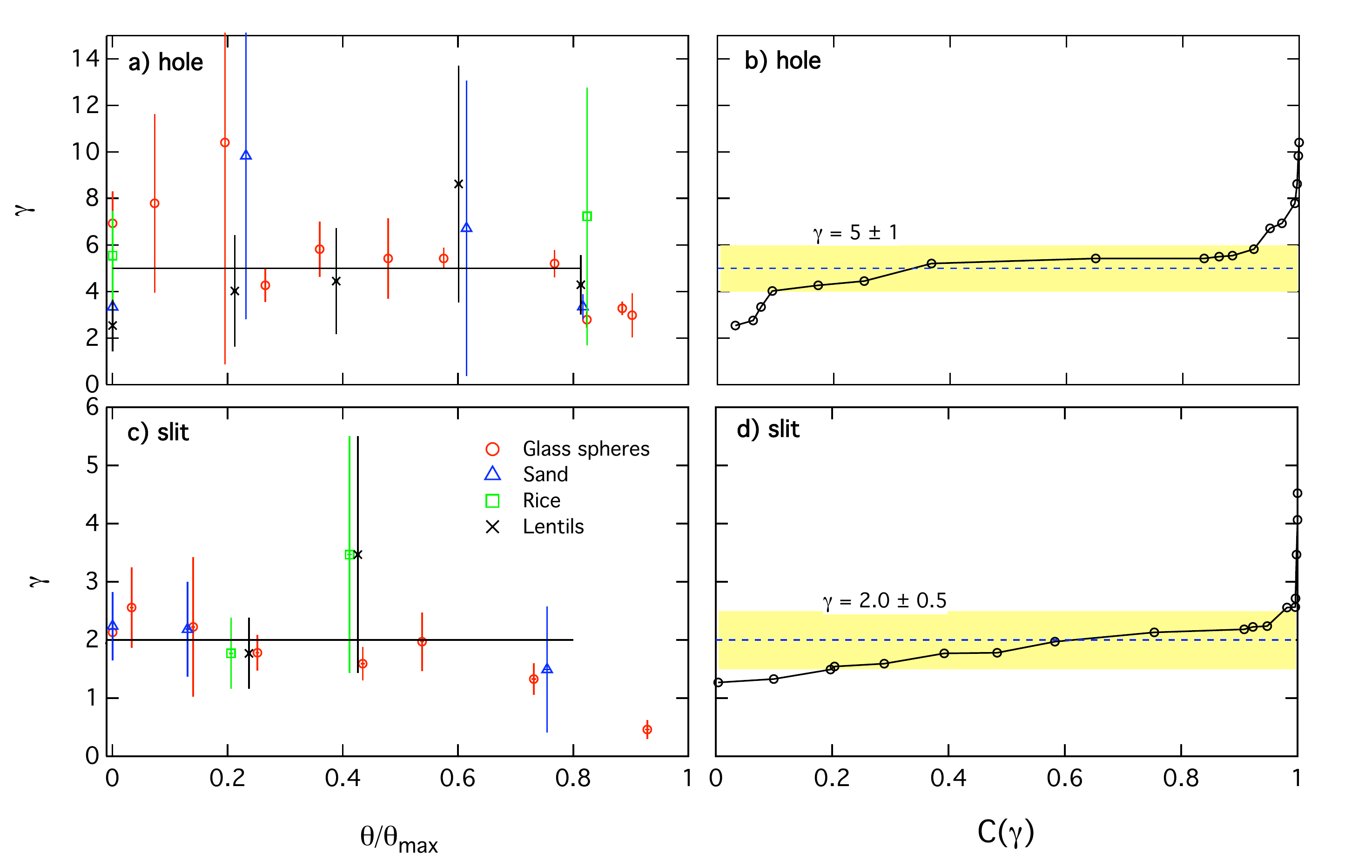}
\caption{(Color online)  Exponents $\gamma$ from fits to Eq.~(\ref{PowerLaw}), versus $\theta/\theta_{max}$, for (a) circular holes and (c) long slits.  The error bars $\delta \gamma$ indicate the fitting confidence intervals. The average values of $\gamma$, weighted by $1/(\delta \gamma)^2$, for $\theta/\theta_{max} <0.8$, and the uncertainty in the means, are $\gamma = 5.1\pm0.3$ for circular holes and $\gamma = 1.9\pm0.1$ for long slits.  The cumulative distributions $C(\gamma)$, with $\Delta C(\gamma)$ increments proportional to $1/(\delta\gamma)^2$, are shown in (b) and (d) for holes and slits, respectively.  There, eighty percent of the weight lies within the ranges $5\pm1$ and $2.0\pm0.1$, as indicated by the yellow shading.  This provides a conservative estimate of the exponents and their uncertainties. 
}
\label{Power}
\end{figure*}

\begin{figure}
\includegraphics[width=3in]{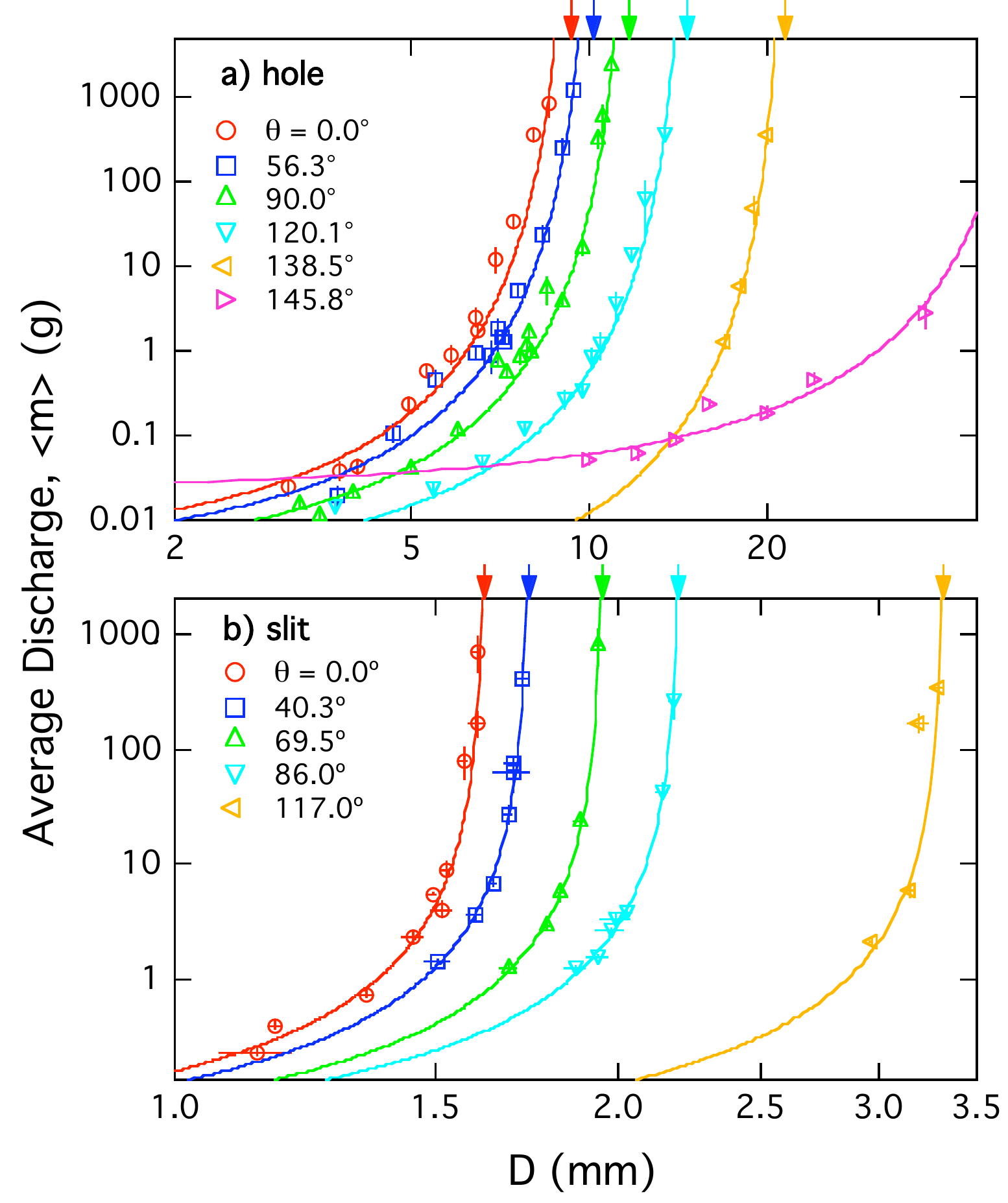}
\caption{(Color online) Average mass discharged versus aperture size $D$ for glass spheres clogging in (a) circular holes and (b) long rectangular slits.  Fits to the diverging power-law of Eq.~(\ref{PowerLaw}) are overlaid, with the exponents fixed to $\gamma = 5$ for circular holes and 2 for rectangular slits, respectively.  Critical aperture sizes from these fits are indicated by arrows.}
\label{MassDdGlass}
\end{figure}

We now determine the location of the clogging transition by using the measurements of the average discharged mass $\langle m \rangle$ for different values of the opening size $D$. Fig.~\ref{MassDZeroDeg} shows how $\langle m \rangle$ grows with aperture size for the various materials, all at $\theta = 0$. For comparison, we have scaled $\langle m \rangle$ by $\rho_b A d$, where $A$ is the aperture area, $\rho_b$ is the material bulk density, and $d$ is the long axis of the grains (see Table~\ref{Materials}). As such, the $y$-axis represents, in units of $d$, the typical height of a column of grains over the aperture that is discharged before a clog forms.   As seen in the figure, the average discharged mass $\langle m \rangle$  grows with $D$, and does so faster than exponential in $D$ or $D^2$ (indicated by the dashed curves).  Instead, the behavior can be well described as a power-law of $D_c-D$, with $\langle m \rangle$ diverging at a critical aperture size $D_c$:
\begin{equation}
	\langle m\rangle = \frac{F}{\left(D_c-D\right)^{\gamma}},
\label{PowerLaw}
\end{equation}
where $F$, $D_c$, and $\gamma$ are fitting parameters~\cite{Zuriguel05, Mankoc09, Janda2D08, To05}. This contrasts with two-dimensional hopper discharge, where the average discharged mass or duration can be fit well by Eq.~(\ref{PowerLaw}), but can equally well be described as growing exponentially with $D$ \cite{Tang09,Tang12} or $D^2$ \cite{To05,Janda2D08}. These forms fail to be generally consistent with all the materials shown in Fig.~\ref{MassDZeroDeg}, although all forms fit the rice data.

For our system, we can determine the value of $D_c$ by fitting the data to Eq.~(\ref{PowerLaw}).  Data points far from the divergence are sometimes excluded, so that all the remaining points above some cutoff are well-fit to the power law form without noticeable systematics in the residuals.  These fits are overlaid as the solid curves in Fig.~\ref{MassDZeroDeg}.  With three free fitting parameters, the fit does not converge for the rice at $\theta = 0$.  This difficulty occasionally occurs for other materials at various values of $\theta$ as well.  Therefore, we next describe our method for determining the best value of $\gamma$ to use for repeating these fits and finding final values for $D_c(\theta)$.

To determine the values of $\gamma$, we first fit the data for each material and $\theta$ to Eq.~(\ref{PowerLaw}), allowing $\gamma$ to float as a free fitting parameter.  Those fitted values of $\gamma$ for which the fits converge are shown in Figs.~\ref{Power}a,c.  The error bars $\delta \gamma$ indicate the fitting confidence intervals. The exponents are scattered over a range of about $\pm 2$ for holes and $\pm 4$ for slits.  There may be a downward trend at very high tilt angles, but for $\theta/\theta_{max}$ less than about 0.8 the scatter appears random.  The weighted average values of $\gamma$ over that range is $5.1 \pm 0.3$ for holes and $1.9 \pm 0.1$ for slits.  The cumulative distributions of $\gamma$ values, $C(\gamma)$, with $\delta C(\gamma)$ increments taken in proportion to $1/(\delta\gamma)^2$, are plotted in Figs.~\ref{Power}b,d.  These show that eighty percent of the weight for the two geometries is covered by $\gamma=5\pm1$ for holes and $\gamma=2.0\pm0.5$.  So we take these as our final best exponent values.   Based on these results, we now repeat the fits to $\langle m\rangle \propto 1/(D_c-D)^\gamma$ using fixed values of $\gamma=5$ and $\gamma=2$ for the holes and slits, respectively. This gives good fits to all the materials and values of $\theta$ for which we have data.  For example, Fig.~\ref{MassDdGlass} shows $\langle m\rangle$ versus $D$ for selected values of $\theta$ for the glass spheres, overlaid with fits to Eq.~(\ref{PowerLaw}). This plot is typical for all the angles and materials studied.

As a final remark for comparison, for three-dimensional hoppers $\gamma = 6.9 \pm 0.2$ was found in Ref.~\cite{Zuriguel05} while $\gamma = 7.6 \pm 0.5$ and $\gamma = 8.6 \pm 0.2$ were found in Ref.~\cite{Mankoc09} for vibrated and non-vibrated hoppers, respectively.  For a two-dimensional hopper, it was reported that $\langle m \rangle$ versus $D$ could be effectively described by both an exponential function of $D^2$ and as a power-law with $\gamma = 11.2$~\cite{To05} and $\gamma = 12.7 \pm 0.1$~\cite{Janda2D08}.

\section{Clogging Phase Diagram}

\begin{figure}
\includegraphics[width=3in]{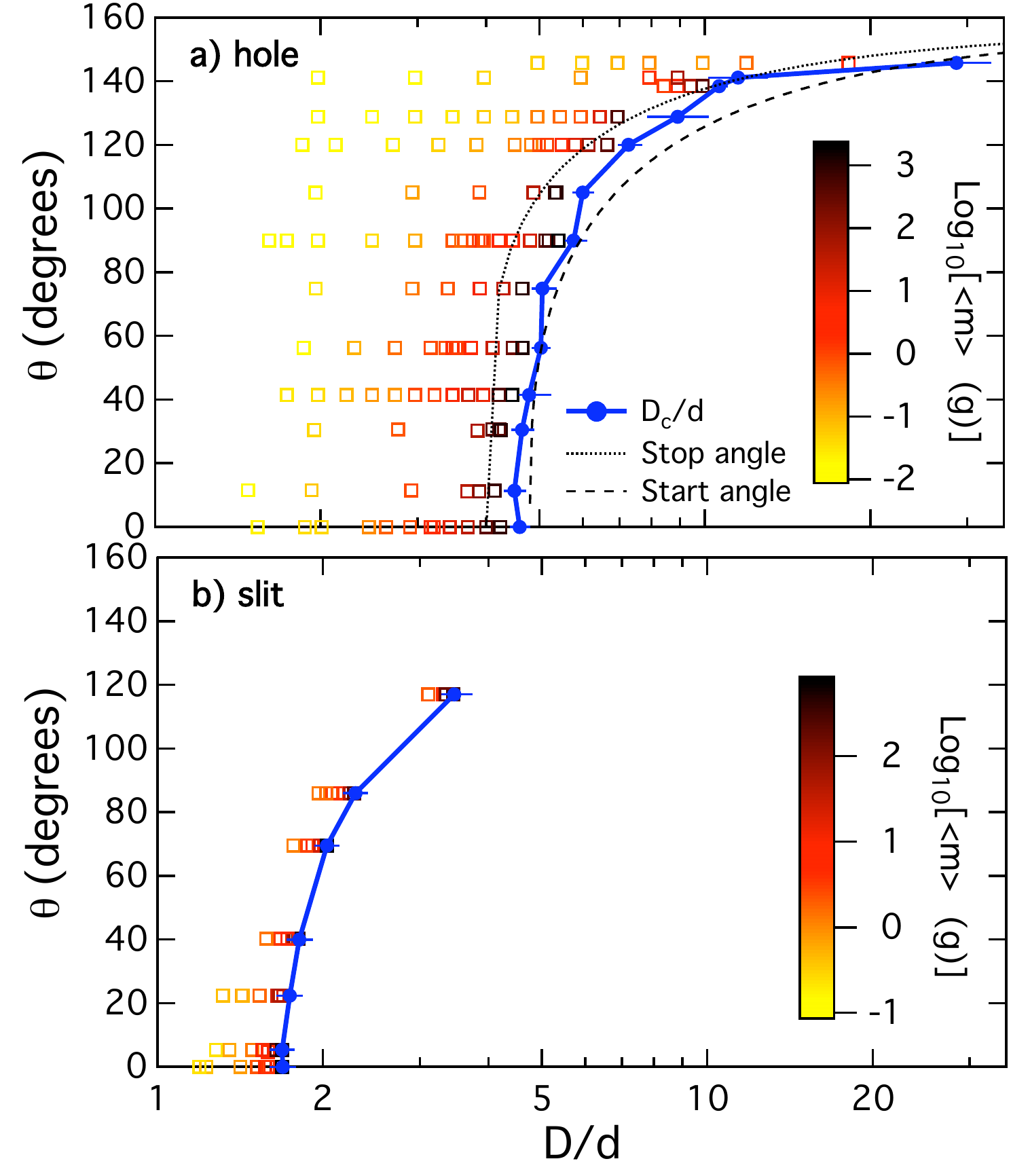}
\caption{(Color online) Clogging phase diagram for glass spheres discharged from (a) circular holes and (b) rectangular slits. The squares, shaded by the average discharged mass $\langle m\rangle$, indicate locations in parameter space where data was collected.  The solid circles show the location of the critical hole diameters or slit widths $D_c$, as determined in Fig.~\ref{MassDdGlass}.  These values of $D_c$ describe a well-defined transition between the clogging and freely-flowing regimes. As indicated by the dotted and dashed lines, respectively, the transition for the circular hole is near the angles where the flow has been observed to spontaneously start and stop during continuous tilting of the hopper \cite{Sheldon}.  The horizontal error bars on $D_c/d$ represent uncertainty from the the power-law fits and the range of acceptable $\gamma$ values.}
\label{TransitionGlass}
\end{figure}

\begin{figure}
\includegraphics[width=3in]{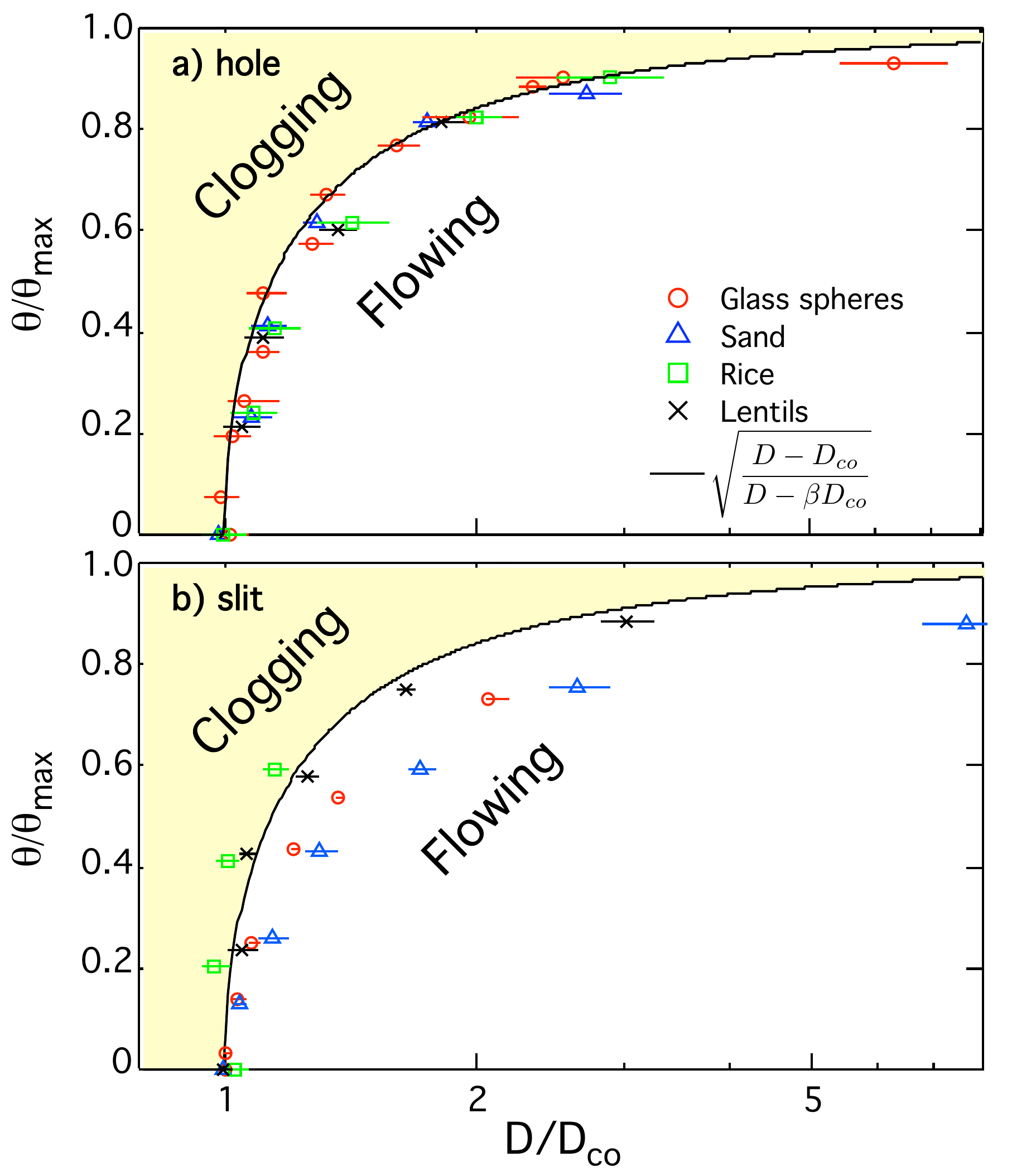}
\caption{(Color online) Clogging transition curves for all materials tested.  We scale the tilt angles $\theta$ by the maximum possible tilt angle for the materials $\theta_{\rm max} = \pi-\theta_r$. We also normalize the critical hole diameters and slit widths $D_c$ by dividing by $D_{co}$, the value of $D_c$ at $\theta/\theta_{max} = 0$ (see Table~\ref{Materials} for values of $\theta_r$ and $D_{co}/d$ for the various materials).  In (a) the transition occurs at the same location in parameter space for all materials.  In (b), however, the location of the transition depends on the material.  An empirical fitting function that captures the behavior for holes is specified in the legend of (a), and is included for comparison in (b); the fitting parameter is $\beta=0.56 \pm 0.02$}
\label{AllTransitions}
\end{figure}

We now map out clogging phase diagrams, which specify whether a given material flows freely or is susceptible to clogging, as a function of aperture size and tilt angle.  Specifically, we follow the previous section in using Eq.~(\ref{PowerLaw}) to fit for the critical hole sizes $D_c$ at which the discharged mass diverges for a range of different tilt angles between 0 and $\pi -\theta_r$.  For glass spheres, the results are collected in Fig.~\ref{TransitionGlass} for both (a) holes and (b) slits.  This figure displays a field of data points where, for a given $\{D,\theta\}$, the value of the average discharged mass is indicated by the shading.  For a given tilt angle, the color darkens as $D$ increases and a solid circle indicates the extrapolated divergence at $D_c$ found by fits to Eq.~(\ref{PowerLaw}).  The locus of $D_c$ values are joined by solid line segments and thus serve to separate the $D-\theta$ parameter space into two regions where clogging does and does not occur.  One can imagine other clogging phase diagrams, where axes could be added to account for the effects of e.g. vibration or other driving forces, but here we focus only on clogging in the $D-\theta$ plane.

The qualitative shape of the clogging phase diagrams in Fig.~\ref{TransitionGlass} is the same for both circular holes and narrow slits, since tilting makes a system more susceptible to clogging.  For both, the transition rises steeply since $D_c$ is nearly constant for small tilt angles.  For larger tilt angles, greater than about ninety degrees, $D_c$ increases rapidly and, intuitively, diverges as $\theta \rightarrow \pi-\theta_r$.  The shape of the transition for circular holes may be compared with the earlier measured from Ref.~\cite{Sheldon} based on start and stop angles.  As seen in Fig.~\ref{TransitionGlass}a, the locus of critical hole sizes matches quite nicely with the locus of start angles.  While this reinforces the validity of Ref.~\cite{Sheldon}, more importantly it shows that the $D-\theta$ clogging phase diagram for a specific material and aperture geometry may be confidently characterized in terms of start angles, which are far easier to measure than critical hole sizes.

Next we investigate how clogging phase diagrams such as Fig.~\ref{TransitionGlass} are affected by the geometry of the grains.  As above, the locus of critical hole sizes $D_c$ are found at a set of tilt angles, now for sand, rice, and lentils.  The various grain types have different sizes and different angles of repose, which affect the location of the locus of clogging transitions in the $D-\theta$ plane.  Therefore, to scale this out and facilitate comparison, we normalize the aperture size by $D_{co}$, the critical hole diameter or slit width at $\theta=0$, and we normalize the tilt angle by $\theta_{\rm max}=\pi-\theta_r$.  Values of $D_{co}$ are given in Table~\ref{Materials}.   The value for $D_{co}$ is of order three times the Beverloo cut-off length $kD$; however, the exact connection between these quantities and grain parameters is unclear.  The resulting scaled clogging phase diagrams for the four grain types are collected in Fig.~\ref{AllTransitions} for (a) holes and (b) slits.  Remarkably, we find a very good collapse of the transition data for all four grain types in Fig.~\ref{AllTransitions}a for circular holes.  This suggests that the clogging phase diagram is universal, independent of grain type.  The shape is satisfactorily described by fit to the empirical form 
\begin{equation}
	\theta/\theta_{\rm max} = Ê\sqrt{ D-D_{co} \over D-\beta D_{co} Ê },
\label{HoleTransitionSqrt}
\end{equation}
with $\beta = 0.56 \pm 0.02$.  However for the case of slits, plotting $\theta/\theta_{\rm max}$ vs $D/D_{co}$ as in Fig.~\ref{AllTransitions}b does not cause collapse for the different grain types.  The transition for lentils is close to the Eq.~(\ref{HoleTransitionSqrt}) fit for holes, but the transitions for other grains have different forms.

\begin{figure}
\includegraphics[width=3in]{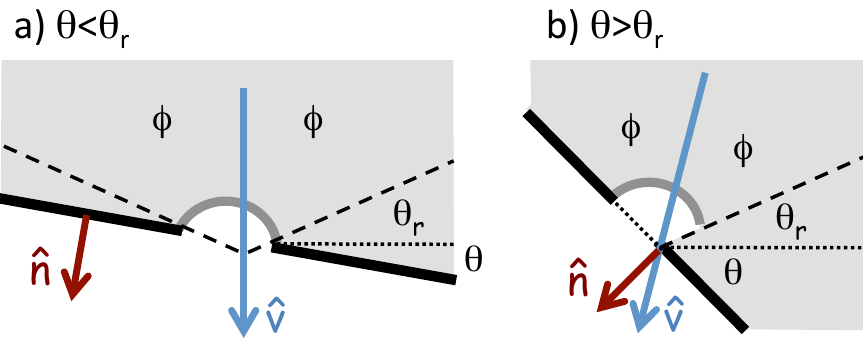}
\caption{(Color online) Geometry of the unit vectors $\hat n$ and $\hat v$, specifying the normal to the plane of the aperture and the average flow direction, respectively, for cases that the tilt angle $\theta$ is (a) less than, and (b) greater than, the angle of repose $\theta_r$.  The average flow direction is defined by bisecting the region where the grains flow, i.e.\ so that the two angles labeled $\phi$ are equal.}
\label{schematic}
\end{figure}

\begin{figure}
\includegraphics[width=3in]{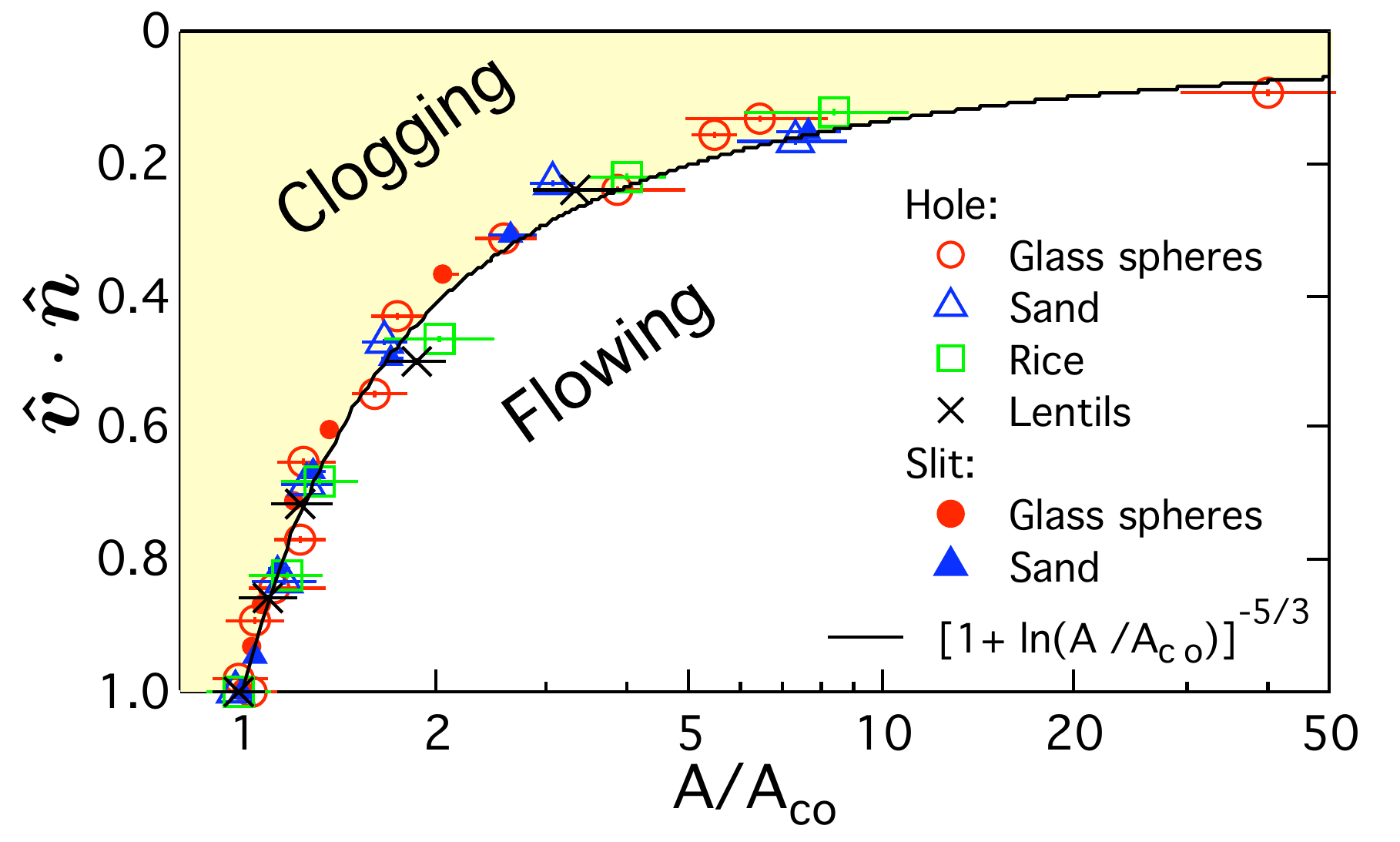}
\caption{(Color online) Clogging transition curves for most materials and {\it both} aperture geometries.  The $x$-axis is the hole area, normalized by the critical hole area at zero tilt angle.  The $y$-axis is the component $\hat v \cdot \hat n$ of the average flow direction normal to the plane of the aperture (see Fig.~\ref{schematic} for illustration defining these unit vectors).  The solid curve is an empirical fitting function; if allowed to float, the exponent is $-1.72\pm0.05$.}
\label{collapse}
\end{figure}

Since there is no theory at present for the shape of the clogging transition curve, we attempt an alternative empirical description to Eq.~(\ref{HoleTransitionSqrt}) based physically on consideration of the direction of the average flow relative to the orientation of the aperture.  Fig.~\ref{schematic} shows a schematic diagram of the system at two different tilt angles.  All grains below the dashed lines inclined at $\theta_r$ above horizontal remain at rest, always.  Only the grains above these lines may flow toward the aperture.  For those grains, the average flow direction is indicated by a unit velocity vector $\hat v$ that bisects the region where flow occurs.  As seen in Fig.~\ref{schematic}, $\hat v$ points straight down for $\theta<\theta_r$ and is inclined for $\theta>\theta_r$.  This defines a projected aperture area as $(A {\hat n})\cdot \hat v$ where $\hat n$ is the unit vector normal to the plane of the aperture, also shown in Fig.~\ref{schematic}.  From the geometry in Fig.~\ref{schematic}, the relevant dot product is computed to be
\begin{equation}
\hat v \cdot \hat n = \cases{
  				 \cos\theta & $\theta \le \theta_r$, \cr
	 			\cos[(\theta+\theta_r)/2] & $\theta \ge \theta_r$. \cr
				}
\label{VdotN}
\end{equation}
For both holes {\it and} slits, the projected area thus decreases the same way from $A$ to zero as $\theta$ increases from 0 to $\theta_{max}=\pi-\theta_r$.  Therefore, we hypothesize that the propensity to clog increases due to the reduction in projected area -- not with respect to gravity but with respect to the average flow direction.  As a test, we replot in Fig.~\ref{collapse} the transition data from Fig.~\ref{AllTransitions}, now as $\hat v \cdot \hat n$ versus $A/A_{co}$ where $A_{co}$ is the critical aperture size at zero tilt.  This causes collapse not just of all transition data for holes, but also for glass spheres and sand data for slits.  For comparison with future theories, one satisfactory empirical fit is to
\begin{equation}
	\hat v \cdot \hat n = [1+ {\rm ln}(A/A_{co})]^{-5/3}
\label{logfit}
\end{equation}
where the left-hand side is given by Eq.~(\ref{VdotN}).   For holes the leading behavior of this form is $\theta\propto\sqrt{D/D_{co}-1}$, the same as for Eq.~(\ref{HoleTransitionSqrt}).  The good collapse for holes and slits means the universality of the clogging transition is greater even than suggested by Fig.~\ref{AllTransitions}.  The only two exceptions are rice and lentils discharged from slits, which are the only instances where both grain and aperture have well-defined axes.  Thus we speculate that orientational ordering of grains with respect to the slit could cause the deviation from Eq.~(\ref{logfit}).

\section{Conclusion}

In this work, we measured the clogging behavior of four different non-cohesive grain types from circular holes and long narrow slits, as a systematic function of both aperture size $D$ and tilt angle $\theta$.  We find that the distribution of discharge events is nearly exponential in all cases, and hence can be well-characterized by the average mass $\langle m \rangle$ discharged between clogging events.  As the aperture size is increased, we find that $\langle m \rangle$ grows as a power-law of $1/(D_c-D)$ and hence diverges at a finite critical aperture size $D_c$.  The exponent depends on the aperture shape:  $\gamma = 5 \pm 1$ for circular holes and $\gamma = 2.0 \pm 0.5$ for slits.  These exponent values are somewhat smaller than those reported in the literature for two-dimensional hoppers \cite{Janda2D08, To05} and three-dimensional hoppers with circular holes \cite{Zuriguel05,Mankoc09}.  However, we are aware neither of prior work reporting $\gamma$ for slits, nor of any models for predicting the values of $\gamma$.

By measuring the critical aperture size for a wide range of tilt angles, we mapped out clogging phase diagrams as a function of aperture size $D$ and hopper tilt angle $\theta$.  In other words, we measured the curves in $\{D,\theta\}$ parameter space that specify whether a particular system is free-flowing, forever, or is susceptible to clogging.  Remarkably, we find that the shape of these curves exhibits a certain universality.  For circular holes, the shape is independent of grain shape when $\theta$ is scaled by $\theta_{max}=\pi-\theta_r$ and $D$ is scaled by the critical diameter at zero tilt.  For long slits and compact grains, the shape is also the same as for the circular holes when the tilt angle is expressed in terms of $\hat v \cdot \hat n$, Eq.~(\ref{VdotN}), and aperture area is scaled by the critical value at zero tilt.  Physically, tilting the sample increases the propensity to clog according to a reduction in the projection of the aperture area against the average flow direction.  This insight, and the striking but unexpected degree of universality in the clogging behavior, now call for a full theoretical explanation.  We believe this is an important challenge, similar in spirit to the notion of deep commonality in the wide classes of jamming transitions.  Another important challenge, which would be of particular benefit to industry, is to extend this whole line of research to grains that are slightly cohesive and hence more susceptible to clogging.  Answers to these questions would provide great insight into the physics of granular materials, as well as to other far-from-equilibrium and disordered systems like vortex pinning and crowds where clogging plays a role.

\begin{acknowledgments}
This work was supported by the NSF through grant DMR-0704147. We thank R.~P.~Behringer for helpful discussions.
\end{acknowledgments}

\bibliography{CloggingRefs}

\begin{thebibliography}{33}
\expandafter\ifx\csname natexlab\endcsname\relax\def\natexlab#1{#1}\fi
\expandafter\ifx\csname bibnamefont\endcsname\relax
  \def\bibnamefont#1{#1}\fi
\expandafter\ifx\csname bibfnamefont\endcsname\relax
  \def\bibfnamefont#1{#1}\fi
\expandafter\ifx\csname citenamefont\endcsname\relax
  \def\citenamefont#1{#1}\fi
\expandafter\ifx\csname url\endcsname\relax
  \def\url#1{\texttt{#1}}\fi
\expandafter\ifx\csname urlprefix\endcsname\relax\def\urlprefix{URL }\fi
\providecommand{\bibinfo}[2]{#2}
\providecommand{\eprint}[2][]{\url{#2}}

\bibitem[{\citenamefont{Jaeger et~al.}(1996)\citenamefont{Jaeger, Nagel, and
  Behringer}}]{JaegerReview}
\bibinfo{author}{\bibfnamefont{H.~M.} \bibnamefont{Jaeger}},
  \bibinfo{author}{\bibfnamefont{S.~R.} \bibnamefont{Nagel}}, \bibnamefont{and}
  \bibinfo{author}{\bibfnamefont{R.~P.} \bibnamefont{Behringer}},
  \bibinfo{journal}{Rev. Mod. Phys.} \textbf{\bibinfo{volume}{68}},
  \bibinfo{pages}{1259} (\bibinfo{year}{1996}).

\bibitem[{\citenamefont{Duran}(2000)}]{DuranBook}
\bibinfo{author}{\bibfnamefont{J.}~\bibnamefont{Duran}},
  \emph{\bibinfo{title}{Sands, powders, and grains: An introduction to the
  physics of granular materials}} (\bibinfo{publisher}{Springer},
  \bibinfo{address}{NY}, \bibinfo{year}{2000}).

\bibitem[{\citenamefont{Committee~on CMMP~2010}(2007)}]{CMMP2010}
\bibinfo{author}{\bibfnamefont{N.~R.~C.} \bibnamefont{Committee~on CMMP~2010},
  \bibfnamefont{Solid State Sciences~Committee}},
  \emph{\bibinfo{title}{Condensed-Matter and Materials Physics: The Science of
  the World around Us}} (\bibinfo{publisher}{The National Academies Press},
  \bibinfo{address}{Washington, DC}, \bibinfo{year}{2007}).

\bibitem[{\citenamefont{To et~al.}(2001)\citenamefont{To, Lai, and Pak}}]{To01}
\bibinfo{author}{\bibfnamefont{K.}~\bibnamefont{To}},
  \bibinfo{author}{\bibfnamefont{P.}~\bibnamefont{Lai}}, \bibnamefont{and}
  \bibinfo{author}{\bibfnamefont{H.~K.} \bibnamefont{Pak}},
  \bibinfo{journal}{Phys. Rev. Lett.} \textbf{\bibinfo{volume}{87}},
  \bibinfo{pages}{71} (\bibinfo{year}{2001}).

\bibitem[{\citenamefont{To and Lai}(2002)}]{To02}
\bibinfo{author}{\bibfnamefont{K.}~\bibnamefont{To}} \bibnamefont{and}
  \bibinfo{author}{\bibfnamefont{P.}~\bibnamefont{Lai}},
  \bibinfo{journal}{Phys. Rev. E} \textbf{\bibinfo{volume}{66}},
  \bibinfo{pages}{011308} (\bibinfo{year}{2002}).

\bibitem[{\citenamefont{Helbing and Johansson}(2006)}]{HelbingPRL}
\bibinfo{author}{\bibfnamefont{D.}~\bibnamefont{Helbing}} \bibnamefont{and}
  \bibinfo{author}{\bibfnamefont{A.}~\bibnamefont{Johansson}},
  \bibinfo{journal}{Phys. Rev. Lett.} \textbf{\bibinfo{volume}{97}},
  \bibinfo{pages}{168001} (\bibinfo{year}{2006}).

\bibitem[{\citenamefont{Chevoir et~al.}(2007)\citenamefont{Chevoir, Gaulard,
  and Roussel}}]{Chevoir07}
\bibinfo{author}{\bibfnamefont{F.}~\bibnamefont{Chevoir}},
  \bibinfo{author}{\bibfnamefont{F.}~\bibnamefont{Gaulard}}, \bibnamefont{and}
  \bibinfo{author}{\bibfnamefont{N.}~\bibnamefont{Roussel}},
  \bibinfo{journal}{Europhys. Lett.} \textbf{\bibinfo{volume}{79}},
  \bibinfo{pages}{14001} (\bibinfo{year}{2007}).

\bibitem[{\citenamefont{Evesque}(2007)}]{Evesque07}
\bibinfo{author}{\bibfnamefont{P.}~\bibnamefont{Evesque}},
  \bibinfo{journal}{Poudres et Grains} \textbf{\bibinfo{volume}{16}},
  \bibinfo{pages}{14} (\bibinfo{year}{2007}).

\bibitem[{\citenamefont{Mort et~al.}(2007)\citenamefont{Mort, Geiger, and
  Wandstrat}}]{Geiger07}
\bibinfo{author}{\bibfnamefont{P.}~\bibnamefont{Mort}},
  \bibinfo{author}{\bibfnamefont{D.}~\bibnamefont{Geiger}}, \bibnamefont{and}
  \bibinfo{author}{\bibfnamefont{M.}~\bibnamefont{Wandstrat}}, in
  \emph{\bibinfo{booktitle}{Conference Proceedings of the 2007 AIChE Annual
  Meeting}} (\bibinfo{address}{Salt Lake City}, \bibinfo{year}{2007}), p.
  \bibinfo{pages}{379e}.

\bibitem[{\citenamefont{Garcimart\'{i}n
  et~al.}(2010)\citenamefont{Garcimart\'{i}n, Zuriguel, Pugnaloni, and
  Janda}}]{Garcimartin10}
\bibinfo{author}{\bibfnamefont{A.}~\bibnamefont{Garcimart\'{i}n}},
  \bibinfo{author}{\bibfnamefont{I.}~\bibnamefont{Zuriguel}},
  \bibinfo{author}{\bibfnamefont{L.~A.} \bibnamefont{Pugnaloni}},
  \bibnamefont{and} \bibinfo{author}{\bibfnamefont{A.}~\bibnamefont{Janda}},
  \bibinfo{journal}{Phys. Rev. E} \textbf{\bibinfo{volume}{82}},
  \bibinfo{pages}{031306} (\bibinfo{year}{2010}).

\bibitem[{\citenamefont{Mersch et~al.}(2010)\citenamefont{Mersch, Lumay,
  Boschini, and Vandewalle}}]{EField}
\bibinfo{author}{\bibfnamefont{E.}~\bibnamefont{Mersch}},
  \bibinfo{author}{\bibfnamefont{G.}~\bibnamefont{Lumay}},
  \bibinfo{author}{\bibfnamefont{F.}~\bibnamefont{Boschini}}, \bibnamefont{and}
  \bibinfo{author}{\bibfnamefont{N.}~\bibnamefont{Vandewalle}},
  \bibinfo{journal}{Phys. Rev. E} \textbf{\bibinfo{volume}{81}},
  \bibinfo{pages}{041309} (\bibinfo{year}{2010}).

\bibitem[{\citenamefont{Zuriguel et~al.}(2011)\citenamefont{Zuriguel, Janda,
  Garcimart\'{\i}n, Lozano, Ar\'evalo, and Maza}}]{Obstacle}
\bibinfo{author}{\bibfnamefont{I.}~\bibnamefont{Zuriguel}},
  \bibinfo{author}{\bibfnamefont{A.}~\bibnamefont{Janda}},
  \bibinfo{author}{\bibfnamefont{A.}~\bibnamefont{Garcimart\'{\i}n}},
  \bibinfo{author}{\bibfnamefont{C.}~\bibnamefont{Lozano}},
  \bibinfo{author}{\bibfnamefont{R.}~\bibnamefont{Ar\'evalo}},
  \bibnamefont{and} \bibinfo{author}{\bibfnamefont{D.}~\bibnamefont{Maza}},
  \bibinfo{journal}{Phys. Rev. Lett.} \textbf{\bibinfo{volume}{107}},
  \bibinfo{pages}{278001} (\bibinfo{year}{2011}).

\bibitem[{\citenamefont{Drescher et~al.}(1995)\citenamefont{Drescher, Waters,
  and Rhoades}}]{Drescher}
\bibinfo{author}{\bibfnamefont{A.}~\bibnamefont{Drescher}},
  \bibinfo{author}{\bibfnamefont{A.~J.} \bibnamefont{Waters}},
  \bibnamefont{and} \bibinfo{author}{\bibfnamefont{C.~A.}
  \bibnamefont{Rhoades}}, \bibinfo{journal}{Powder Technol.}
  \textbf{\bibinfo{volume}{84}}, \bibinfo{pages}{177} (\bibinfo{year}{1995}).

\bibitem[{\citenamefont{Helbing et~al.}(2000)\citenamefont{Helbing, Farkas, and
  Vicsek}}]{HelbingNature}
\bibinfo{author}{\bibfnamefont{D.}~\bibnamefont{Helbing}},
  \bibinfo{author}{\bibfnamefont{I.}~\bibnamefont{Farkas}}, \bibnamefont{and}
  \bibinfo{author}{\bibfnamefont{T.}~\bibnamefont{Vicsek}},
  \bibinfo{journal}{Nature} \textbf{\bibinfo{volume}{407}},
  \bibinfo{pages}{487} (\bibinfo{year}{2000}).

\bibitem[{\citenamefont{Liu and Nagel}(2001)}]{LiuNagelBook}
\bibinfo{editor}{\bibfnamefont{A.~J.} \bibnamefont{Liu}} \bibnamefont{and}
  \bibinfo{editor}{\bibfnamefont{S.~R.} \bibnamefont{Nagel}}, eds.,
  \emph{\bibinfo{title}{Jamming and Rheology: Constrained Dynamics on
  Microscopic and Macroscopic Scales}} (\bibinfo{publisher}{Taylor and
  Francis}, \bibinfo{address}{NY}, \bibinfo{year}{2001}).

\bibitem[{\citenamefont{Liu et~al.}(2011)\citenamefont{Liu, Nagel, van
  Saarloos, and Wyart}}]{LiuNagelChapter}
\bibinfo{author}{\bibfnamefont{A.~J.} \bibnamefont{Liu}},
  \bibinfo{author}{\bibfnamefont{S.~R.} \bibnamefont{Nagel}},
  \bibinfo{author}{\bibfnamefont{W.}~\bibnamefont{van Saarloos}},
  \bibnamefont{and} \bibinfo{author}{\bibfnamefont{M.}~\bibnamefont{Wyart}}, in
  \emph{\bibinfo{booktitle}{Dynamical Heterogeneities in Glasses, Colloids and
  Granular Media}}, edited by
  \bibinfo{editor}{\bibfnamefont{L.}~\bibnamefont{Berthier}},
  \bibinfo{editor}{\bibfnamefont{G.}~\bibnamefont{Biroli}},
  \bibinfo{editor}{\bibfnamefont{J.}~\bibnamefont{Bouchaud}},
  \bibinfo{editor}{\bibfnamefont{L.}~\bibnamefont{Cipelleti}},
  \bibnamefont{and} \bibinfo{editor}{\bibfnamefont{W.}~\bibnamefont{van
  Saarloos}} (\bibinfo{publisher}{Oxford University Press},
  \bibinfo{address}{Oxford}, \bibinfo{year}{2011}), pp.
  \bibinfo{pages}{298--340}.

\bibitem[{\citenamefont{Reichhardt and Reichhardt}(2010)}]{Reichhardt}
\bibinfo{author}{\bibfnamefont{C.~J.~O.} \bibnamefont{Reichhardt}}
  \bibnamefont{and}
  \bibinfo{author}{\bibfnamefont{C.}~\bibnamefont{Reichhardt}},
  \bibinfo{journal}{Phys. Rev. B} \textbf{\bibinfo{volume}{81}},
  \bibinfo{pages}{224516} (\bibinfo{year}{2010}).

\bibitem[{\citenamefont{Reichhardt et~al.}(2012)\citenamefont{Reichhardt,
  Groopman, Nussinov, and Reichhardt}}]{Reichhardt2012}
\bibinfo{author}{\bibfnamefont{C.~J.~O.} \bibnamefont{Reichhardt}},
  \bibinfo{author}{\bibfnamefont{E.}~\bibnamefont{Groopman}},
  \bibinfo{author}{\bibfnamefont{Z.}~\bibnamefont{Nussinov}}, \bibnamefont{and}
  \bibinfo{author}{\bibfnamefont{C.}~\bibnamefont{Reichhardt}},
  \bibinfo{journal}{arXiv:1204.6342v1}  (\bibinfo{year}{2012}).

\bibitem[{\citenamefont{Bug et~al.}(2012)\citenamefont{Bug, Bullard-Sisken,
  Goodrich, Manning, and Liu}}]{GoodrichBug}
\bibinfo{author}{\bibfnamefont{A.~L.~R.} \bibnamefont{Bug}},
  \bibinfo{author}{\bibfnamefont{S.}~\bibnamefont{Bullard-Sisken}},
  \bibinfo{author}{\bibfnamefont{C.~P.} \bibnamefont{Goodrich}},
  \bibinfo{author}{\bibfnamefont{M.~L.} \bibnamefont{Manning}},
  \bibnamefont{and} \bibinfo{author}{\bibfnamefont{A.~J.} \bibnamefont{Liu}},
  \bibinfo{journal}{Bull. Am. Phys. Soc.} \textbf{\bibinfo{volume}{57}},
  \bibinfo{pages}{J53.5} (\bibinfo{year}{2012}).

\bibitem[{\citenamefont{Zuriguel et~al.}(2005)\citenamefont{Zuriguel,
  Garcimart\'{i}n, Maza, Pugnaloni, and Pastor}}]{Zuriguel05}
\bibinfo{author}{\bibfnamefont{I.}~\bibnamefont{Zuriguel}},
  \bibinfo{author}{\bibfnamefont{A.}~\bibnamefont{Garcimart\'{i}n}},
  \bibinfo{author}{\bibfnamefont{D.}~\bibnamefont{Maza}},
  \bibinfo{author}{\bibfnamefont{L.~A.} \bibnamefont{Pugnaloni}},
  \bibnamefont{and} \bibinfo{author}{\bibfnamefont{J.~M.}
  \bibnamefont{Pastor}}, \bibinfo{journal}{Phys. Rev. E}
  \textbf{\bibinfo{volume}{71}}, \bibinfo{pages}{051303}
  (\bibinfo{year}{2005}).

\bibitem[{\citenamefont{Mankoc et~al.}(2009)\citenamefont{Mankoc,
  Garcimart\'{i}n, Zuriguel, Maza, and Pugnaloni}}]{Mankoc09}
\bibinfo{author}{\bibfnamefont{C.}~\bibnamefont{Mankoc}},
  \bibinfo{author}{\bibfnamefont{A.}~\bibnamefont{Garcimart\'{i}n}},
  \bibinfo{author}{\bibfnamefont{I.}~\bibnamefont{Zuriguel}},
  \bibinfo{author}{\bibfnamefont{D.}~\bibnamefont{Maza}}, \bibnamefont{and}
  \bibinfo{author}{\bibfnamefont{L.~A.} \bibnamefont{Pugnaloni}},
  \bibinfo{journal}{Phys. Rev. E} \textbf{\bibinfo{volume}{80}},
  \bibinfo{pages}{011309} (\bibinfo{year}{2009}).

\bibitem[{\citenamefont{To}(2005)}]{To05}
\bibinfo{author}{\bibfnamefont{K.}~\bibnamefont{To}}, \bibinfo{journal}{Phys.
  Rev. E} \textbf{\bibinfo{volume}{71}}, \bibinfo{pages}{060301}
  (\bibinfo{year}{2005}).

\bibitem[{\citenamefont{Janda et~al.}(2008)\citenamefont{Janda, Zuriguel,
  Garcimart\'{i}n, Pugnaloni, and Maza}}]{Janda2D08}
\bibinfo{author}{\bibfnamefont{A.}~\bibnamefont{Janda}},
  \bibinfo{author}{\bibfnamefont{I.}~\bibnamefont{Zuriguel}},
  \bibinfo{author}{\bibfnamefont{A.}~\bibnamefont{Garcimart\'{i}n}},
  \bibinfo{author}{\bibfnamefont{L.~A.} \bibnamefont{Pugnaloni}},
  \bibnamefont{and} \bibinfo{author}{\bibfnamefont{D.}~\bibnamefont{Maza}},
  \bibinfo{journal}{Europhys. Lett.} \textbf{\bibinfo{volume}{84}},
  \bibinfo{pages}{44002} (\bibinfo{year}{2008}).

\bibitem[{\citenamefont{Tang et~al.}(2009)\citenamefont{Tang, Sagdiphour, and
  Behringer}}]{Tang09}
\bibinfo{author}{\bibfnamefont{J.}~\bibnamefont{Tang}},
  \bibinfo{author}{\bibfnamefont{S.}~\bibnamefont{Sagdiphour}},
  \bibnamefont{and} \bibinfo{author}{\bibfnamefont{R.~P.}
  \bibnamefont{Behringer}}, \bibinfo{journal}{AIP Conference Proceedings}
  \textbf{\bibinfo{volume}{1145}}, \bibinfo{pages}{515} (\bibinfo{year}{2009}).

\bibitem[{\citenamefont{Tang et~al.}(2012)\citenamefont{Tang, Behringer, and
  Mort}}]{Tang12}
\bibinfo{author}{\bibfnamefont{J.}~\bibnamefont{Tang}},
  \bibinfo{author}{\bibfnamefont{R.~P.} \bibnamefont{Behringer}},
  \bibnamefont{and} \bibinfo{author}{\bibfnamefont{P.}~\bibnamefont{Mort}},
  \bibinfo{journal}{Bull. Am. Phys. Soc.} \textbf{\bibinfo{volume}{57}},
  \bibinfo{pages}{X53.7} (\bibinfo{year}{2012}).

\bibitem[{\citenamefont{Sheldon and Durian}(2010)}]{Sheldon}
\bibinfo{author}{\bibfnamefont{H.~G.} \bibnamefont{Sheldon}} \bibnamefont{and}
  \bibinfo{author}{\bibfnamefont{D.~J.} \bibnamefont{Durian}},
  \bibinfo{journal}{Granular Matter} \textbf{\bibinfo{volume}{12}},
  \bibinfo{pages}{579} (\bibinfo{year}{2010}).

\bibitem[{\citenamefont{Pouliquen}(1999)}]{Pouliquen}
\bibinfo{author}{\bibfnamefont{O.}~\bibnamefont{Pouliquen}},
  \bibinfo{journal}{Phys. Fluids} \textbf{\bibinfo{volume}{11}},
  \bibinfo{pages}{542} (\bibinfo{year}{1999}).

\bibitem[{\citenamefont{Beverloo et~al.}(1961)\citenamefont{Beverloo, Leniger,
  and van~de Velde}}]{Beverloo}
\bibinfo{author}{\bibfnamefont{W.~A.} \bibnamefont{Beverloo}},
  \bibinfo{author}{\bibfnamefont{H.~A.} \bibnamefont{Leniger}},
  \bibnamefont{and} \bibinfo{author}{\bibfnamefont{J.}~\bibnamefont{van~de
  Velde}}, \bibinfo{journal}{Chem. Eng. Sci.} \textbf{\bibinfo{volume}{15}},
  \bibinfo{pages}{260} (\bibinfo{year}{1961}).

\bibitem[{\citenamefont{Nedderman et~al.}(1982)\citenamefont{Nedderman, Tuzun,
  Savage, and Houlsby}}]{NeddermanSavage}
\bibinfo{author}{\bibfnamefont{R.~M.} \bibnamefont{Nedderman}},
  \bibinfo{author}{\bibfnamefont{U.}~\bibnamefont{Tuzun}},
  \bibinfo{author}{\bibfnamefont{S.~B.} \bibnamefont{Savage}},
  \bibnamefont{and} \bibinfo{author}{\bibfnamefont{G.~T.}
  \bibnamefont{Houlsby}}, \bibinfo{journal}{Chem. Eng. Sci.}
  \textbf{\bibinfo{volume}{37}}, \bibinfo{pages}{1597} (\bibinfo{year}{1982}).

\bibitem[{\citenamefont{Anand et~al.}(2008)\citenamefont{Anand, Curtis,
  Wassgren, Hancock, and Ketterhagen}}]{Anand08}
\bibinfo{author}{\bibfnamefont{A.}~\bibnamefont{Anand}},
  \bibinfo{author}{\bibfnamefont{J.~S.} \bibnamefont{Curtis}},
  \bibinfo{author}{\bibfnamefont{C.~R.} \bibnamefont{Wassgren}},
  \bibinfo{author}{\bibfnamefont{B.~C.} \bibnamefont{Hancock}},
  \bibnamefont{and} \bibinfo{author}{\bibfnamefont{W.~R.}
  \bibnamefont{Ketterhagen}}, \bibinfo{journal}{Chem. Eng. Sci.}
  \textbf{\bibinfo{volume}{63}}, \bibinfo{pages}{5821} (\bibinfo{year}{2008}).

\bibitem[{\citenamefont{Zuriguel et~al.}(2003)\citenamefont{Zuriguel,
  Pugnaloni, Garcimart\'{i}n, and Maza}}]{Zuriguel03}
\bibinfo{author}{\bibfnamefont{I.}~\bibnamefont{Zuriguel}},
  \bibinfo{author}{\bibfnamefont{L.~A.} \bibnamefont{Pugnaloni}},
  \bibinfo{author}{\bibfnamefont{A.}~\bibnamefont{Garcimart\'{i}n}},
  \bibnamefont{and} \bibinfo{author}{\bibfnamefont{D.}~\bibnamefont{Maza}},
  \bibinfo{journal}{Phys. Rev. E} \textbf{\bibinfo{volume}{68}},
  \bibinfo{pages}{030301} (\bibinfo{year}{2003}).

\bibitem[{\citenamefont{Janda et~al.}(2009)\citenamefont{Janda, Maza,
  Garcimart\'{\i}n, Kolb, Lanuza, and Clement}}]{Janda09}
\bibinfo{author}{\bibfnamefont{A.}~\bibnamefont{Janda}},
  \bibinfo{author}{\bibfnamefont{D.}~\bibnamefont{Maza}},
  \bibinfo{author}{\bibfnamefont{A.}~\bibnamefont{Garcimart\'{\i}n}},
  \bibinfo{author}{\bibfnamefont{E.}~\bibnamefont{Kolb}},
  \bibinfo{author}{\bibfnamefont{J.}~\bibnamefont{Lanuza}}, \bibnamefont{and}
  \bibinfo{author}{\bibfnamefont{E.}~\bibnamefont{Clement}},
  \bibinfo{journal}{Europhys. Lett.} \textbf{\bibinfo{volume}{87}},
  \bibinfo{pages}{24002} (\bibinfo{year}{2009}).

\bibitem[{\citenamefont{Saraf and Franklin}(2011)}]{Saraf}
\bibinfo{author}{\bibfnamefont{S.}~\bibnamefont{Saraf}} \bibnamefont{and}
  \bibinfo{author}{\bibfnamefont{S.~V.} \bibnamefont{Franklin}},
  \bibinfo{journal}{Phys. Rev. E} \textbf{\bibinfo{volume}{83}},
  \bibinfo{pages}{030301} (\bibinfo{year}{2011}).

\end{thebibliography}
\end{document}